\documentclass[fleqn,usenatbib]{mnras}

\usepackage{newtxtext,newtxmath}

\usepackage[T1]{fontenc}

\DeclareRobustCommand{\VAN}[3]{#2}
\let\VANthebibliography\thebibliography
\def\thebibliography{\DeclareRobustCommand{\VAN}[3]{##3}\VANthebibliography}

\usepackage{multirow}
\usepackage{graphicx}	
\usepackage{amsmath}	
\usepackage{xcolor}

\newcommand{\revised}[1]{{#1}}

\defcitealias{Mertens_2020}{M20}
\defcitealias{Acharya_2024}{A24}

\title[ML+GPR on LOFAR EoR 21-cm data]{Revised LOFAR upper limits on the 21-cm signal power spectrum at $\mathbf{z\approx9.1}$ using Machine Learning and Gaussian Process Regression}

\author[A. Acharya et al.]{Anshuman Acharya$^{1}$\thanks{E-mail: anshuman@mpa-garching.mpg.de (AA)},
Florent Mertens$^{2}$,
Benedetta Ciardi$^{1}$,
Raghunath Ghara$^{3,4}$,
L\'eon V. E. Koopmans$^{5}$,
\newauthor and Saleem Zaroubi$^{5,6}$
\\
$^{1}$Max-Planck-Institut für Astrophysik, Garching 85748, Germany \\
$^{2}$LERMA, Observatoire de Paris, PSL Research University, CNRS, Sorbonne Universit\'{e}, F-75014 Paris, France \\
$^{3}$Haverford College, 370 Lancaster Ave, Haverford PA, 19041, USA \\
$^{4}$Center for Particle Cosmology, Department of Physics and Astronomy, University of Pennsylvania, Philadelphia, PA 19104, USA \\
$^{5}$Kapteyn Astronomical Institute, University of Groningen, PO Box 800, 9700AV Groningen, The Netherlands \\
$^{6}$Astrophysics Research Centre of the Open University of Israel, Ra’anana 4353701, Israel \\
}

\date{Accepted XXX. Received YYY; in original form ZZZ}

\pubyear{20XX}

\begin{document}
\label{firstpage}
\pagerange{\pageref{firstpage}--\pageref{lastpage}}
\maketitle

\begin{abstract}

\noindent The use of Gaussian Process Regression (GPR) for foregrounds mitigation in data collected by the LOw-Frequency ARray (LOFAR) to measure the high-redshift 21-cm signal power spectrum has been shown to have issues of signal loss when the 21-cm signal covariance is misestimated. To address this problem, we have recently introduced covariance kernels obtained by using a Machine Learning based Variational Auto-Encoder (VAE) algorithm in combination with simulations of the 21-cm signal. In this work, we apply this framework to 141 hours ($\approx 10$ nights) of LOFAR data at $z \approx 9.1$, and report revised upper limits of the 21-cm signal power spectrum. Overall, we agree with past results reporting a 2-$\sigma$ upper limit of $\Delta^2_{21} < (80)^2~\rm mK^2$ at $k = 0.075~h~\rm Mpc^{-1}$. Further, the VAE-based kernel has a smaller correlation with the systematic excess noise, and the overall GPR-based approach is shown to be a good model for the data. Assuming an accurate bias correction for the excess noise, we report a 2-$\sigma$ upper limit of $\Delta^2_{21} < (25)^2~\rm mK^2$ at $k = 0.075~h~\rm Mpc^{-1}$. However, we still caution to take the more conservative approach to jointly report the upper limits of the excess noise and the 21-cm signal components.
\end{abstract}



\section{Introduction}\label{sec:intro}

The eras between redshifts $z \sim 30$ to $z \sim 5$, when the first astrophysical objects formed and started to emit photons that ionized the neutral hydrogen in the intergalactic medium (IGM), are known as the Cosmic Dawn and the Epoch of Reionization (EoR). The brightness fluctuation of the 21-cm line of neutral hydrogen as observed in emission or absorption against the Cosmic Microwave Background (CMB), is one of the most promising probes to study this redshift range \citep[e.g.][]{Field_1959,Madau_1997,Shaver_1999,Tozzi_2000,Ciardi_2003,Zaroubi_2013}. Even a statistical upper limit estimate of the strength of such fluctuations would allow us to constrain models of structure formation in the early Universe \citep{Ghara2020, Mondal_2020,Greig_2021,Greig_2021lofar,Abdurashidova_2022}. 

While multiple interferometric low-frequency radio telescopes have been designed to search for this signal (e.g. PAPER\footnote{Precision Array to Probe EoR, \url{http://eor.berkeley.edu}}, MWA\footnote{Murchison Widefield Array, \url{http://www.mwatelescope.org}}, HERA\footnote{Hydrogen Epoch of Reionization Array, \url{https://reionization.org/}}, NenuFAR\footnote{New Extension in Nançay Upgrading LOFAR, \url{https://nenufar.obs-nancay.fr/en}} and the upcoming SKA\footnote{Square Kilometre Array, \url{https://www.skao.int/en}}), in this work we focus on LOFAR\footnote{Low-Frequency Array, \url{http://www.lofar.org}}. The most recent and stringent 2-$\sigma$ upper limits on the power of the 21-cm signal from 141 hours ($\approx$10 nights) of observation with LOFAR at $z$=9.1, were provided by \citet[][hereafter M20]{Mertens_2020} as $\Delta^2_{21} (k = 0.075~h~\rm Mpc^{-1}) < (73)^2~\rm mK^2$. 

As the 21-cm signal is seen through foregrounds that are several orders of magnitude stronger than the signal itself, one of the challenges faced by all 21-cm signal experiments is to develop an accurate modeling and removal of the foregrounds. The LOFAR EoR Key Science Project (KSP) team employs Gaussian Process Regression (GPR) to model different components of the observed data, as described by \citet{Mertens2018}, \citet{Gehlot_2019} and \citet{Hothi_2021}. \revised{In this method, each contributing component of the data is represented by a Gaussian Process, which completely depends on the mean and the covariance between the range of frequencies considered. This covariance can thus be represented as a matrix, which is also referred to as a ``covariance kernel''.} However, given the choice of normalization and bias correction in the power-spectra estimation used by \citetalias{Mertens_2020}, \citet{Kern_2021} noted the dependence of the results on the choice of the covariance kernel used for the 21-cm signal so that any misestimation of the kernel could lead to spurious signal gain/loss. This in turn would affect the astrophysical interpretation of the estimated 21-cm signal power spectrum. 

To improve the set of 21-cm signal covariance kernels, \citet{Mertens_2024} introduced the concept of a learned covariance kernel, using a Variational Auto-Encoder (VAE) algorithm applied to 21-cm signal simulations. This was employed by \citet[][hereafter A24]{Acharya_2024} after training on {\sc Grizzly} simulations of reionization \citep{Ghara_2015,Ghara_2018,Ghara2020}, and tested against various mock datasets. In this work, we apply the VAE-based covariance kernels trained at $z=9.16$ by \citetalias{Acharya_2024} to the same $\approx$10 nights of data used by \citetalias{Mertens_2020}, and compare its performance with that of the Exponential Matern-class function based kernel employed by \citetalias{Mertens_2020}. 

In Section~\ref{sec:methods}, we briefly highlight the covariance kernels used for the different components of the dataset; in Section~\ref{sec:results}, we present two cases of recovery, and  compare the obtained upper limits to those presented in  \citetalias{Mertens_2020}. In  Section~\ref{sec:discuss}, we discuss the quality of our results by comparing the residuals on applying our model to the data, and also the limitations of the method. Finally, in Section~\ref{sec:conc}, we give our conclusions.

\section{Methodology}\label{sec:methods}

\begin{table}
\centering
\caption{
Comparison of hyperparameters obtained in \citet{Mertens_2020} and in this work. From left to right the columns refer to the hyperparameters used, the results from \citetalias{Mertens_2020}, the results obtained in this work, and the difference between the two. As the covariance kernels used for the 21-cm signal component are not the same, we just provide the values of the hyperparameters obtained in each case. Lastly, $\sigma_{\rm noise}^2$ is $74~\times 10^{3}~\rm mK^2$ (see \citetalias{Mertens_2020}).}
\begin{tabular}{llll}
\hline 
Parameter & \citetalias{Mertens_2020} & This work & Difference \\
\hline
\vspace{0.2em}
$l_{\rm sky}$ & $47.5^{+3.1}_{-2.8}$ & $38.6^{+2.6}_{-2.6}$ & 2.3$\sigma$ \\
\vspace{0.2em}
$\sigma_{\rm sky}^2/\sigma_{\rm noise}^2$ & $611^{+22}_{-19}$ & 530$\pm$18 & 3.0$\sigma$  \\
\vspace{0.2em}
$l_{\rm mix}$ & $2.97^{+0.09}_{-0.08}$ & 3.05$^{+0.08}_{-0.08}$ & 0.7$\sigma$ \\
\vspace{0.2em}
$\sigma_{\rm mix}^2/\sigma_{\rm noise}^2$ & $50.4^{+2.1}_{-1.9}$ & $48.8^{+2.1}_{-2.0}$ & 0.6$\sigma$ \\
\vspace{0.2em}
$l_{\rm ex}$ & $0.26^{+0.01}_{-0.01}$ & 0.32$^{+0.01}_{-0.01}$ & 4.2$\sigma$ \\
\vspace{0.2em}
$\sigma_{\rm ex}^2/\sigma_{\rm noise}^2$ & $2.18^{+0.09}_{-0.14}$ & $2.82^{+0.10}_{-0.10}$ & 4.2$\sigma$ \\
\hline
\vspace{0.2em}
\multirow{3}{*}{$f_{\rm 21}$} & $\eta_{\rm 21} = 1/2$ & $x_1 = 0.49^{+1.00}_{-1.00}$, & \\
& $l_{\rm 21} > 0.73$ & $x_2 = -0.34^{+0.97}_{-0.97}$ & --\\
& $\frac{\sigma_{\rm 21}^2}{\sigma_{\rm noise}^2} < 0.77$ & $\frac{\sigma_{\rm 21}^2}{\sigma_{\rm noise}^2} = 0.01^{+0.09}_{-0.01}$ & \\
\hline
\end{tabular}
\label{table:params}
\end{table}

Gaussian Process Regression \citep[GPR;][]{RasmussenBook,Aigrain_2023} has been used to model radio data in frequency space as noisy observations of the form $\mathbf{y} = \mathbf{f(x)} + \epsilon$, with $\epsilon$ Gaussian noise vector having variance $\sigma_{\rm noise}^2$ \citep[see e.g. \citetalias{Acharya_2024}, \citetalias{Mertens_2020} and][]{Munshi_2024}. The vector $\mathbf{f(x)}$ can be split into the 21-cm signal ($\mathbf{f}_{\rm 21}$) and foregrounds. To the latter contribute an intrinsic sky component ($\mathbf{f}_{\rm sky}$) and mode-mixing contaminants ($\mathbf{f}_{\rm mix}$). Further, \citetalias{Mertens_2020} identified an additional term for bias correction, defining it as an ``excess noise''  component ($\mathbf{f}_{\rm ex}$), which corrects for additional systematic noise after the subtraction of foregrounds. \revised{While the exact source of this excess is yet to be identified, various possibilities have been discussed in Section 6.2 of \citetalias{Mertens_2020}, and the LOFAR EoR KSP team is currently working on testing each one of them.} The overall datacube ${\bf y}$ can be expressed as: \begin{equation}\label{eq:splitdata}
{\bf y} = \mathbf{f}_{\rm sky}(\mathbf{x}) + \mathbf{f}_{\rm mix}(\mathbf{x}) + \mathbf{f}_{\rm ex}(\mathbf{x}) + \mathbf{f}_{\rm 21}(\mathbf{x}) + \epsilon .
\end{equation}

For a set of points {\bf x} (independent parameters in frequency), \revised{for each value $x$, $f(x)$ given by a Gaussian Process $\mathcal{GP}(m,\kappa)$ is fully defined by its mean $m$ and covariance matrix $\kappa$. The vector \textbf{f(x)} is then fully defined by its mean vector \textbf{m} and covariance matrix vector \textbf{K}. The joint distribution for all random variables that share the desired covariance properties is a normal distribution vector  ($\mathcal{N}$) which can be represented as}:
\begin{equation}\label{eq:gp}
    \mathbf{f(x)} \sim \mathcal{N} ( \mathbf{m(x)}, \mathbf{K(x, x)} ) \, .
\end{equation} 
Here, the matrix \textbf{K} gives the covariance between the function values at any two points, and can be written as $K_{ij} = \kappa(x_i,x_j,\phi) + \delta_{ij}\sigma_{i, \rm noise}^2$, where $\kappa(x_i,x_j,\phi)$ can be optimised by the choice of hyperparameters represented by $\phi$, $\delta_{ij}$ is the Kronecker-delta function and $\sigma_{i, \rm noise}^2$ is the noise at $x_i$. 

If we assume that the various components are uncorrelated, we can utilise the additive property of matrices. Thus, the covariance kernels of the right-hand side of Equation~\ref{eq:splitdata} can be represented as a single covariance kernel $K$ given as:
\begin{equation}\label{eq:kerneq}
    \mathbf{K = K_{\rm sky} + K_{\rm mix} + K_{\rm noise} + K_{\rm ex} + K_{\rm 21}} \, .
\end{equation}
$\mathbf{K}_{\rm sky}$, $\mathbf{K}_{\rm mix}$, and $\mathbf{K}_{\rm ex}$ are modelled using the best-fit Matern-class functions \citep[][]{Stein1999} as done by \citetalias{Mertens_2020} and \citetalias{Acharya_2024}: 
\begin{equation}\label{eq:Maternfunc}
    k_{\rm Matern}(r) = \sigma^2 \frac{2^{1-\eta}}{\Gamma (\eta)} \left( \frac{\sqrt{2\eta }r}{l} \right)^{\eta} \kappa_{\eta} \left( \frac{\sqrt{2\eta}r}{l} \right),
\end{equation} 
with $\eta_{\rm sky} = +\infty$, $\eta_{\rm mix} = 3/2$ and $\eta_{\rm ex} = 5/2$. Additionally, $l$ is the coherence-scale hyperparameter, with its associated variance given by $\sigma^2$, $r$ is the absolute difference between the frequencies of two sub-bands, $\kappa_{\eta}$ is the modified Bessel function of the second kind, and $\Gamma$ is the Gamma-function. \revised{$\eta_{\rm sky},~\eta_{\rm mix},~\rm and~\eta_{\rm ex}$ were obtained by \citetalias{Mertens_2020} by assuming different values of the hyperparameter $\eta$, and finding the one that maximised the marginal likelihood (or the ``evidence''). This was done by calculating the analytical integral over \textbf{f}, which is the log-marginal-likelihood (LML, see Section 2.3 in \citet{Mertens2018}). While \citetalias{Mertens_2020} used a gradient-descent-based optimization algorithm
for maximising the LML, \citetalias{Acharya_2024}, instead, introduced an MCMC sampling based approach \citep{Foreman_Mackey_2013} to estimate the hyperparameters by sampling their posterior distributions. This additionally provides a measure of the uncertainty on the hyperparameters.}

The noise is modelled based on the Stokes-V visibility difference, according to the methodology laid out in \citetalias{Mertens_2020}. Further, for the 21-cm signal we use the VAE-based kernel at $z=9.16$ built by \citetalias{Acharya_2024} by training on \textsc{Grizzly} simulations. This VAE-based kernel is fully defined using two hyperparameters $x_1$ and $x_2$ with uninformed flat priors in linear space in the range [-10, 10], and an associated variance. \revised{The kernel has been trained on a training set of $\approx 1500$ simulations with 4 independently variable parameters. While more hyperparameters could be employed, they do not provide any significant improvement, and thus we do not use them to avoid overfitting. \citetalias{Acharya_2024} also used an additional testing set of $\approx 150$ simulations and found a recovery error of $\lesssim 1\%$ for wave-modes of $k < 0.43 h\rm Mpc^{-1}$.} In this work, the performance of the VAE-based kernel is compared against the results obtained in \citetalias{Mertens_2020} when using a Matern-class function with $\eta_{\rm 21} = 1/2$.

\begin{figure}
\centering
\includegraphics[width=\columnwidth,keepaspectratio]{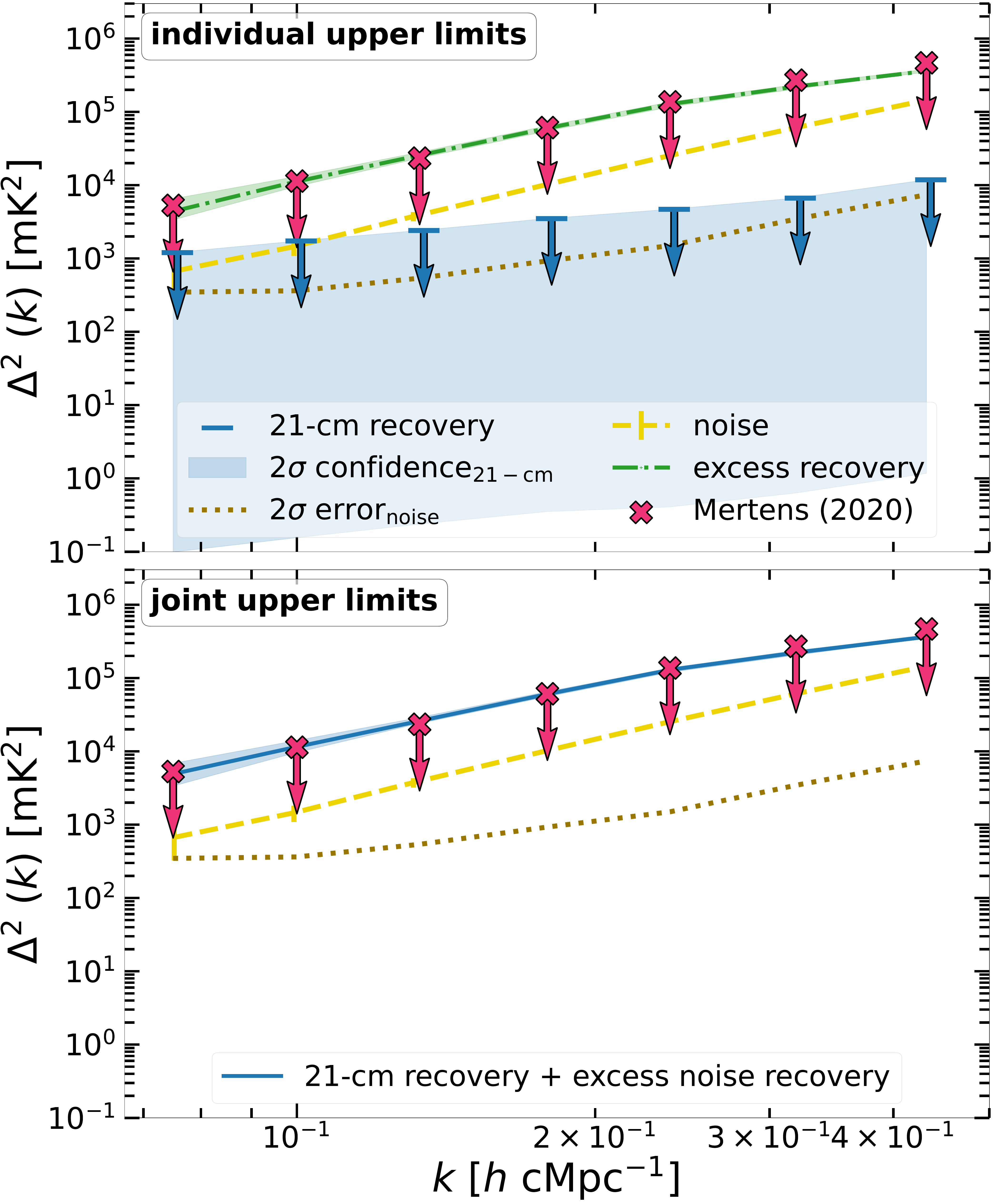}
\caption{Upper limits on the power spectrum of the 21-cm signal from 10 nights of LOFAR observational data at $z \approx 9.1$. \textbf{Top:} Individual upper limits case. The upper limits on $\Delta^2_{21}$ (blue dashes with downward arrows) are obtained using the VAE-based kernel separately from the excess noise (dashed-dotted green line). The 2-$\sigma$ confidence of the recovery by the VAE-based kernel is shaded in blue, and noted to effectively extend down to zero. We also plot the upper limits obtained by \citetalias{Mertens_2020} (crosses with downward arrows), the noise (dashed yellow line) and the 2-$\sigma$ error on the noise (dotted brown).
\textbf{Bottom:} Joint upper limits case. The upper limits on the power spectrum of the 21-cm signal and excess noise are recovered together (solid blue line and shaded region), as done in \citetalias{Mertens_2020}. The other lines and symbols have the same meaning as in the top panel.
} 
\label{fig_mainresult}
\end{figure}

\section{Results}\label{sec:results}

Following \citetalias{Acharya_2024}, we recover the hyperparameters of the various signal components by applying GPR to the data using an MCMC approach \citep{Foreman_Mackey_2013}. We adopt the same broad flat priors for the variances, and uniform priors for the coherence-scale parameters (see Table 1 and Section 2.5 of \citetalias{Acharya_2024}). The recovered values are listed in Table~\ref{table:params}, where we compare them to those obtained by \citetalias{Mertens_2020}. 
We find that the hyperparameters for the $f_{\rm sky}$ and $f_{\rm ex}$ components differ from the \citetalias{Mertens_2020} estimates by about 2-4$\sigma$, while those for $f_{\rm mix}$ are in good agreement. \revised{This is expected with a more accurate 21-cm kernel, as it reduces the strength of the correlation of hyperparameters of the different components. This allows for better characterisation of the different components in the overall data. We discuss the details below after analysing each component's recovered power spectra.}

To have a better understanding of the results quantified above, in Figure~\ref{fig_mainresult}, we show the recovered upper limits on the power spectra in two cases. In the first one (top panel), we present the excess noise component  (dashed-dotted green line) and the 21-cm signal (blue dashes with downward arrows) separately, as done for the mock datasets in \citetalias{Acharya_2024}, and refer to this as the \textit{individual upper limits case}. We additionally show the 2-$\sigma$ confidence interval on the recovery of the VAE-based kernel, the overall noise (yellow dashed line), and the 2-$\sigma$ error on the noise (brown dotted line). We compare our recovery with the results from \citetalias{Mertens_2020} shown with magenta crosses. We note that while the upper limits are higher than the 2-$\sigma$ error on the noise, the values approach similar values at the higher $k$ bins (see values listed in Table~\ref{table:recovery}). Further, it is still possible that some fraction of the 21-cm signal may be assigned to the excess noise, and vice-versa. Thus, we also show a second case (bottom panel), which takes a more conservative approach of including the excess noise in the upper limits along with the 21-cm signal (solid blue line), as done in \citetalias{Mertens_2020}. We refer to this as the \textit{joint upper limits case}.

We note that, in the individual case, the upper limits on the 21-cm signal power spectrum are significantly deeper than those of the combined power spectrum of the excess noise and 21-cm signal found by \citetalias{Mertens_2020}.

Because we depict just the extracted 21-cm signal component, we assume that the bias correction for the excess noise is accurate, with minimal loss of the 21-cm signal. \revised{This is an idealised scenario, where the excess noise component is accurately described by the covariance kernel used for it. To verify this, we need a better understanding of the source of the excess noise component. Nevertheless, we report results with our current best choice of the kernel as identified by \citetalias{Mertens_2020}, to test the extent of improvement possible with the inclusion of the VAE-based kernel for the 21-cm signal. However, in Figure~\ref{fig_mainresult}, the 2-$\sigma$ confidence interval on the lower side extends to $< 1~\rm mK^2$, and thus is effectively zero. Thus, this confirms} that this is just an upper limit and not a detection.

\begin{table*}
\centering
\caption{
Comparing the results of \citet{Mertens_2020} with those of our joint and individual upper limits of the excess noise component and 21-cm signal power spectra. 
The first column refers to the $k$ bins, and the following to the recovered 21-cm signal power ($\Delta^2_{21}$) and its corresponding 2-$\sigma$ upper limit ($\Delta^2_{\rm 21, UL}$) for \citetalias{Mertens_2020}, the joint upper limits (bottom panel in Figure~\ref{fig_mainresult}) and the individual upper limits (top panel in Figure~\ref{fig_mainresult}) applied in this work. The last column shows the 2-$\sigma$ error on the noise (2-$\sigma$ error$_{\rm noise}$).}
\begin{tabular}{llllllll}
\hline 
$k$ & \multicolumn{2}{c}{\citetalias{Mertens_2020} ($f_{\rm ex} + f_{\rm 21}$)}& \multicolumn{2}{c}{joint upper limits ($f_{\rm ex} + f_{\rm 21}$)} & \multicolumn{2}{c}{individual upper limits ($f_{\rm 21}$)} & 2-$\sigma$ error$_{\rm noise}$\\
 & $\Delta^2_{21}$ & $\Delta^2_{\rm 21, UL}$ & $\Delta^2_{21}$ & $\Delta^2_{\rm 21, UL}$ & $\Delta^2_{21}$ & $\Delta^2_{\rm 21, UL}$ & $2\Delta^2_{\rm noise, err}$ \\
$(h~\rm cMpc^{-1})$ & (mK$^2$) & (mK$^2$) & (mK$^2$) & (mK$^2$) & (mK$^2$) & (mK$^2$) & (mK$^2$) \\
\hline
0.075 & (58.96)$^2$ & (72.86)$^2$ & (71.33)$^2$ & (79.68)$^2$ & (6.58)$^2$ & (24.55)$^2$ & (18.63)$^2$ \\
0.100 & (95.21)$^2$ & (106.65)$^2$ & (107.05)$^2$ & (115.77)$^2$ & (8.56)$^2$ & (29.43)$^2$ & (19.05)$^2$ \\
0.133 & (142.17)$^2$ & (153.00)$^2$ & (158.65)$^2$ & (167.76)$^2$ & (10.48)$^2$ & (37.29)$^2$ & (23.22)$^2$ \\
0.179 & (235.80)$^2$ & (246.92)$^2$ & (245.42)$^2$ & (261.40)$^2$ & (14.40)$^2$ & (43.92)$^2$ & (30.47)$^2$ \\
0.238 & (358.95)$^2$ & (370.18)$^2$ & (360.08)$^2$ & (372.28)$^2$ & (16.29)$^2$ & (56.32)$^2$ & (38.70)$^2$ \\
0.319 & (505.26)$^2$ & (520.33)$^2$ & (472.49)$^2$ & (490.22)$^2$ & (19.95)$^2$ & (73.79)$^2$ & (58.72)$^2$ \\
0.432 & (664.23)$^2$ & (683.20)$^2$ & (601.37)$^2$ & (626.81)$^2$ & (26.61)$^2$ & (99.19)$^2$ & (86.01)$^2$ \\
\hline
\end{tabular}
\label{table:recovery}
\end{table*}

When comparing the joint upper limits cases, we find marginally higher values than those found by \citetalias{Mertens_2020}\ for $k \lesssim 0.2~h~\rm cMpc^{-1}$, and marginally lower for $0.2~h~{\rm cMpc^{-1}} \lesssim k \lesssim 0.5~h~\rm cMpc^{-1}$. This is caused by the power of the excess noise component being mildly higher in the lower $k$-bins, and mildly lower in the higher $k$-bins, compared to the results of \citetalias{Mertens_2020}. This shows up as a 4.2$\sigma$ difference of the hyperparameters $l_{\rm ex}$ and $\sigma_{\rm ex}^2/\sigma_{\rm noise}^2$ from the results of \citetalias{Mertens_2020}. A closer inspection shows that this occurs because a small fraction of the intrinsic sky component is assigned to the excess noise component at lower $k$-bins, and vice-versa for the higher $k$-bins, \revised{due to the usage of a more accurate 21-cm kernel. While the difference in the recovered power spectrum for the foregrounds is not significant, the higher contributions at small scales for the excess noise, could allow a better characterisation of the same.}

The exact values of the recovered median power spectrum \revised{of the 21-cm signal} and their upper limits in each $k$-bin are listed in Table~\ref{table:recovery} for the joint and individual upper limits case discussed in this work, as well by \citetalias{Mertens_2020}. We also show the 2-$\sigma$ error of the noise \revised{(which represents the maximum sensitivity achievable)} to compare against the obtained upper limits. We can summarize the results as follows: 
\begin{itemize}
    \item The excess noise is confirmed to be the dominant signal component after sky-model and residual foregrounds subtraction, with its recovered power spectrum being about an order of magnitude stronger than the noise.
    \item While the hyperparameters for the excess noise component differ by 4.2$\sigma$ from the results of \citetalias{Mertens_2020}, the recovered power spectrum and 2-$\sigma$ upper limits show only a minor deviation.
    \item In the case of individual upper limits, the upper limits of the 21-cm signal power spectrum lie above the 2-$\sigma$ error on the noise, and thus are not noise-dominated. However, this is not a detection, because the lower limit on the recovery by the VAE-based kernel is consistent with zero.
\end{itemize}

We conclude that with these data, performing a bias correction for the excess noise provides an individual upper limit on the 21-cm signal power spectrum that is higher than the 2-$\sigma$ error on the noise. However, the upper limits approach the 2-$\sigma$ error on the noise at higher $k$ bins. Thus, lowering the noise threshold with longer observations could allow a greater separation between the upper limits and the noise threshold, and eventually also a detection, provided no correlation between the excess noise and the 21-cm signal is present. For this, the LOFAR EoR KSP team has $\approx 100$ nights of data to analyse and is currently in the process of doing so. We discuss the correlations of the different components and the limitations of the overall model in the subsequent section.

\section{Discussion}\label{sec:discuss}

\begin{figure}
\centering
\includegraphics[width=\columnwidth,keepaspectratio]{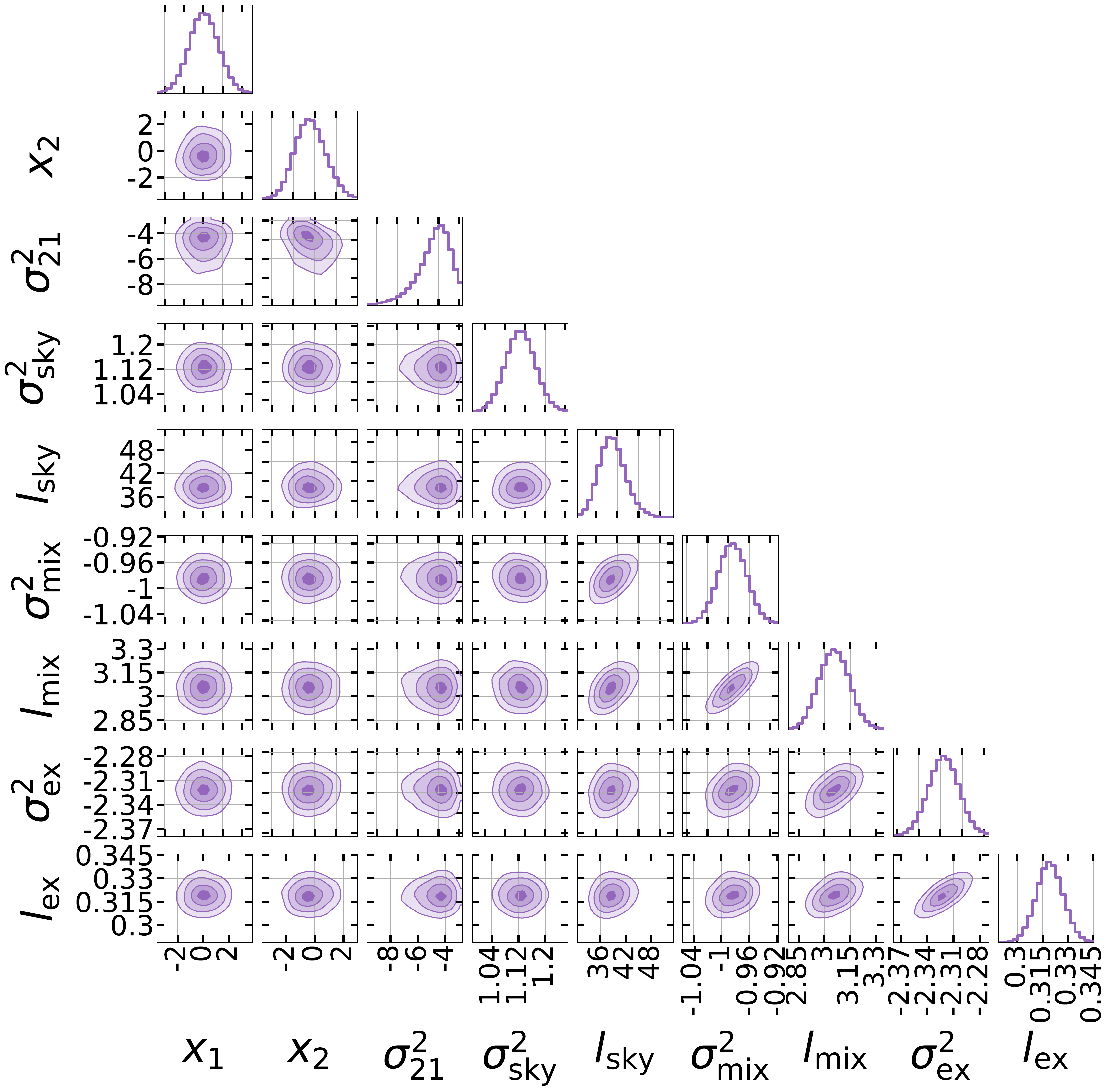}
\caption{Posterior probability distributions of the GPR model hyperparameters used: $x_1$, $x_2$, the variances $\sigma^2_{21}$, $\sigma_{\rm sky}^2$, $\sigma_{\rm mix}^2$ and $\sigma_{\rm ex}^2$, and the coherence-scales $l_{\rm sky}$, $l_{\rm mix}$ and $l_{\rm ex}$. The purple contours show the 68\%, 95\% and 99.7\% confidence intervals and the diagonal plots refer to the individual posterior distributions of each of the hyperparameters listed. Note that the hyperparameters of each component are largely uncorrelated with those of other components.} 
\label{fig_cp}
\end{figure}

\begin{figure}
\centering
\includegraphics[width=\columnwidth,keepaspectratio]{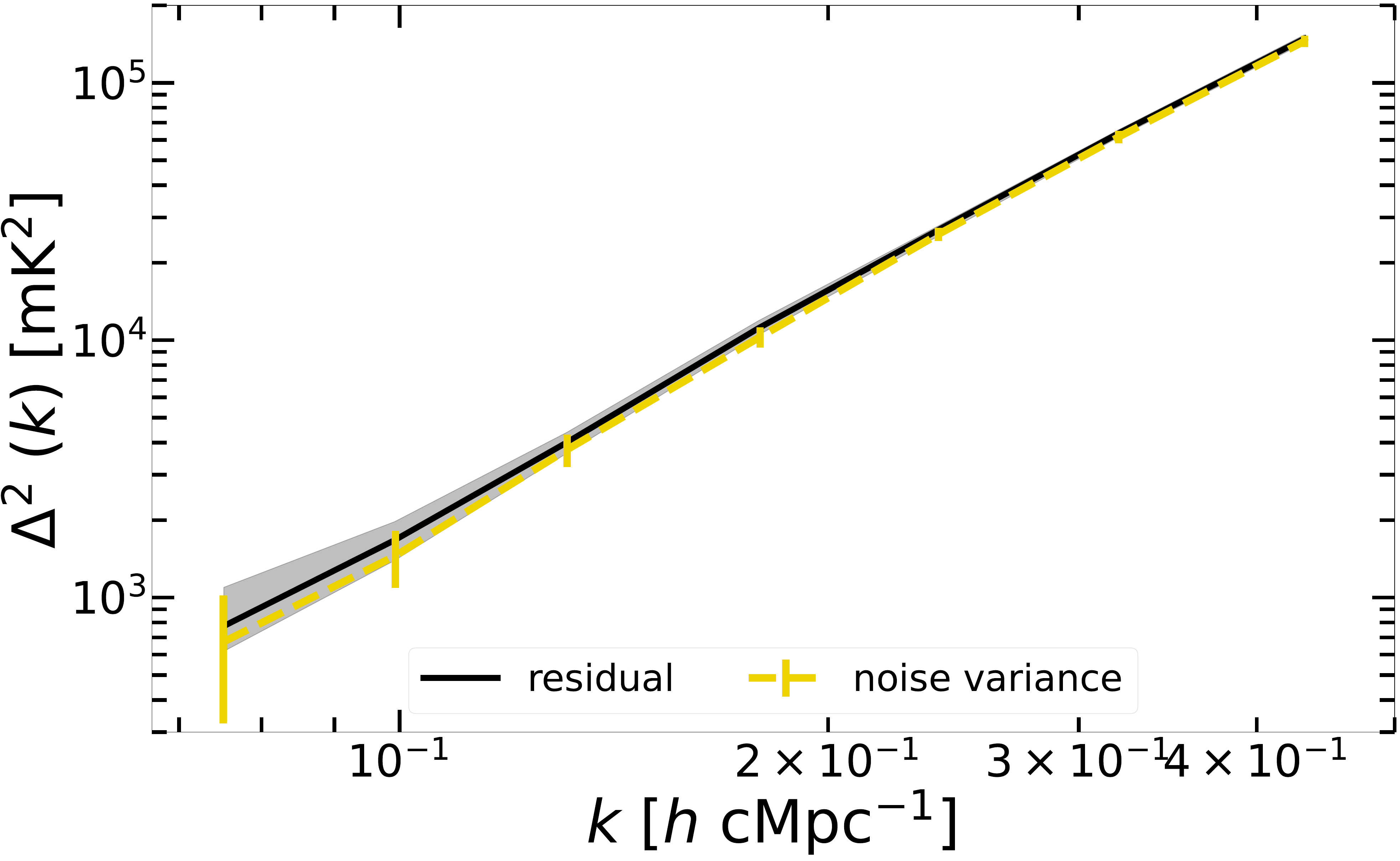}
\caption{Residual (black solid line) obtained by subtracting the full GPR model from the data, and its corresponding 2-$\sigma$ confidence interval (grey shaded region).
The noise is shown as a yellow dashed line with 2-$\sigma$ error bars. We note an excellent agreement between them.} 
\label{fig_noisecompare}
\end{figure}

The individual upper limits are greater than the 2-$\sigma$ error on the noise, and thus is not noise dominated. However, it is still possible that the VAE-based kernels provide an incomplete model of the data. In this case, we would be biased towards cases where the model is enough for the 21-cm signal, and the remainder of the data is modelled with the excess and foregrounds components as defined above. For example, better characterisation of the excess noise may indeed lead to broader models for it. In our current model, what we can explore however, is the degeneracy between the VAE-based kernel and the excess noise kernel. A decrease in the correlation between them would indicate that at least the models we use for them are not degenerate. For example, the difference in the intrinsic sky and excess noise component hyperparameters with respect to \citetalias{Mertens_2020} could be due to a decrease in correlation of the hyperparameters for these components by using the VAE-based kernel for the 21-cm signal. To have a clearer picture of this, it is important to investigate whether the hyperparameters of the different components in our overall model are correlated and whether the overall model is a complete picture of the data. 

To assess this, in Figure~\ref{fig_cp}, we show the corner plot of the hyperparameters, where the purple contours are the 68\%, 95\% and 99.7\% confidence intervals. We see that hyperparameters of each component are largely uncorrelated with those of other components. In particular, we note that the 21-cm signal hyperparameters are completely uncorrelated from those of the excess noise. This provides greater confidence in the 2-$\sigma$ confidence intervals obtained in the individual upper limits case.

Furthermore, to assess the performance of our model, we plot the residual obtained by subtracting the data cube of the model from the observational data cube in Figure~\ref{fig_noisecompare}, and compare its power spectrum (black solid) to that of the noise data cube (yellow dashed). We find excellent agreement between them within the 2-$\sigma$ error on the noise. To numerically quantify this, we calculate the $\chi^2$ of the residual versus the noise power spectrum, and find this to be $\approx 6.12$. With 7 $k$ bins, this gives a reduced $\chi^2 \approx 0.87$, thus confirming the performance of the model. However, it is still possible that some minor leakage can occur between the different signal components. Thus, better modelling of the excess noise remains necessary to improve the bias correction as more data is added. Furthermore, for an eventual detection, reduction of the noise component through longer observations is also needed.

\section{Conclusions}\label{sec:conc}

To improve the modelling of the 21-cm signal covariance kernel, \citet{Mertens_2024} introduced a Machine Learning based method that uses a Variational Auto-Encoder (VAE) based algorithm. This was trained on {\sc Grizzly} simulations \citep{Ghara_2015,Ghara_2018,Ghara2020} of the 21-cm signal and tested against mock datasets by \citet{Acharya_2024}. In this work, we use the VAE-based kernel trained at $z=9.16$ to obtain an updated upper limit on the 21-cm signal from 141 hours of LOFAR data ($\approx 10$ nights) at $z \approx 9.1$. We compare our results to past efforts, which used analytic functions instead of Machine Learning \citep{Mertens_2020}, and find that they are consistent with errors, although  the new upper limits are slightly increased at $k \lesssim 0.2~h~\rm cMpc^{-1}$, and mildly decreased for $0.2~h~{\rm cMpc^{-1}}\lesssim k \lesssim 0.5~h~\rm cMpc^{-1}$. Overall, we report a 2-$\sigma$ upper limit of $\Delta^2_{21} < (80)^2~\rm mK^2$ at $k = 0.075~h~\rm Mpc^{-1}$ in the conservative case, when jointly recovering the 21-cm signal and excess noise components. 

We also investigate the case in which the upper limits of the 21-cm signal are evaluated including a bias correction for the excess noise component. In this case, we report a 2-$\sigma$ upper limit of $\Delta^2_{21} < (25)^2~\rm mK^2$ at $k = 0.075~h~\rm Mpc^{-1}$. We confirm that the VAE kernel hyperparameters are uncorrelated from the excess component ones, thus providing confidence for the estimated upper limits. This is strengthened by the fact that the residuals obtained by subtracting the full model from the data are consistent with the noise with $0.87 \sigma$. Although the bias correction is promising, we still caution against using the bias-corrected upper limits, as better modelling of the excess noise remains necessary to improve confidence on its separation from the 21-cm signal, as the data improves and the signal to noise increases. In future work, the LOFAR EoR KSP team will be improving the characterisation of the excess noise component and its corresponding covariance kernel, and assess whether an excess noise bias correction can reliably be applied as this work suggests it can. \revised{As demonstrated here, this approach would substantially reduce the current upper limits on the 21-cm signal, although this should still be considered as an upper limit and not a detection for the given data, as it is already reaching the sensitivity limit due to the noise, given by the 2-$\sigma$ error on the noise. This improvement could offer enhanced astrophysical constraints compared to \citet{Ghara2020}, allowing for the rejection of a significantly greater number of cold IGM scenarios and achieving stricter constraints on the sources of X-ray heating. This shall be carried out in future work.} Additionally, the team will make use of VAE-based covariance kernels as developed in \citet{Acharya_2024} across multiple redshifts with data from the LOFAR telescope.

\section*{Acknowledgements}
AA thanks the EoR research group at MPA for helpful discussions. RG acknowledges support from the Kaufman Foundation (Gift no. GF01364). LVEK acknowledges the financial support from the European Research Council (ERC) under the European Union's Horizon 2020 research and innovation programme (Grant agreement No. 884760, `CoDEX'). SZ thanks Alexander von Humboldt Foundation for the Humboldt Research award and MPA hospitality. \revised{Lastly, we would like to thank the referee for their helpful comments that contributed to the improvement of this paper.}

\section*{Data Availability}
The VAE kernels for $z = 9.16$ can be shared on reasonable request. The {\sc ps\_eor} package used to run GPR and ML-GPR can be found at \url{https://gitlab.com/flomertens/ps_eor}. The LOFAR data analysed in this work can be shared on reasonable request.



\bibliographystyle{mnras}
\bibliography{mlgpr_10nights} 



\bsp	
\label{lastpage}
\end{document}